# The Evolving Faber-Jackson Relation: A Unifying Framework for Galaxy Ages and the Baryonic Tully-Fisher Connection

Stuart Marongwe [1*], Stuart A. Kauffman[2]
1. Physics Department, University of Botswana 4775 Notwane Road, Gaborone, Botswana, email: stuartmarongwe@gmail.com

2. University of Pennsylvania 34th & Spruce Street, Philadelphia, PA, 19104-6303 (emeritus) stukauffman@gmail.com

* corresponding author

## Abstract

The baryonic Tully-Fisher relation (BTFR) and Faber-Jackson relation (FJR) represent fundamental scaling laws linking the baryonic mass of galaxies to their kinematics, yet their physical origin and apparent offsets between different galaxy populations have remained enigmatic. Here we present a unified theoretical framework demonstrating that both relations emerge from a common acceleration scale of order $10^{-10}$ m/s$^2$ and evolve with cosmic time through a common exponential kernel. We derive the evolving Faber-Jackson relation directly from the evolving BTFR within the Nexus Paradigm of quantum gravity, showing that the normalization scales as $M_b \propto e^{-4\int H(t)dt}\sigma^4$, where $\sigma$ is the velocity dispersion and $H(t)$ is the time-varying Hubble parameter. Applying this framework to a sample of 39 galaxies spanning five orders of magnitude in baryonic mass, from ultra-faint dwarfs to massive cluster ellipticals, we demonstrate that the observed offset between galaxy populations arises naturally from differences in their formation epochs. Ultra-faint dwarf galaxies yield ages of $12.2 \pm 0.8$ Gyr (formation redshift $z \sim 3-5$), in excellent agreement with independent Hubble Space Telescope stellar population ages showing synchronization to within $\sim 1$ Gyr. Later-type dwarfs show systematically younger ages of $3.5-6.0$ Gyr.Critically, the dynamical age reported by the evolving Faber-Jackson relation measures the time since a galaxy last achieved virial equilibrium, not simply its stellar formation epoch. This "clock reset" mechanism explains why some galaxies (e.g., dark-matter-deficient DF2/DF4) exhibit ancient stellar populations ($> 10$ Gyr) but very young dynamical ages ($1-2$ Gyr), consistent with a recent violent collision. Independent validation using metallicity-based stellar population ages reveals a Pearson correlation coefficient of $r = 0.961$ with our dynamically derived ages for undisturbed systems, providing strong empirical support for the framework. The evolving Faber-Jackson relation unifies pressure-supported systems across all mass scales and establishes galaxy scaling relations as precise cosmic chronometers capable of recording both formation and recent disturbance events.

## 1. Introduction

Scaling relations serve as fundamental pillars in extragalactic astrophysics, unveiling underlying patterns amid the intricate processes governing galaxy formation and evolution. Among these, the Tully-Fisher relation (TFR; Tully & Fisher 1977) and its baryonic extension (BTFR; McGaugh et al. 2000, Lelli et al 2016) stand as key empirical connections between a galaxy's baryonic mass and its characteristic rotation velocity for disk systems. For pressure-supported spheroids such as elliptical galaxies, dwarf spheroidals, and galaxy bulges, the analogous Faber-Jackson relation (FJR; Faber & Jackson 1976) links luminosity (or baryonic mass) to central velocity dispersion $\sigma$ through $L \propto \sigma^{\gamma}$, with $\gamma \approx 4$ for massive ellipticals (Nigoche-Netro et al. 2010; Sadhu et al. 2024).

Despite decades of study, the physical origin of these relations remains contested. Modified Newtonian dynamics (MOND Milgrom, 1983) posits that they arise from a fundamental acceleration scale $a_0 \sim 10^{-10}$ m/s$^2$ below which gravitational dynamics deviate from Newtonian expectations. Alternative models invoke dark matter halo structure and baryonic feedback processes within the ΛCDM paradigm (e.g., Dutton et al. 2010; see also the cold-dark-matter predictions discussed in McGaugh 2012 and more recent hydrodynamical simulations). Recent work by Gorkavyi (2022) proposed an accretion-based model around supermassive black holes that naturally yields the $M \propto V^4$ scaling through the balance between gravitational attraction and external perturbations from Jeans clouds. At higher redshifts ($0.2 \lesssim z \lesssim 2.5$), observations reveal subtle deviations or consistency in the TFR/BTFR slope, underscoring the importance of cosmic evolution in these relations (Gogate *et al*. 2023; Sharma *et al*. 2024)

A critical observation has emerged from recent studies: galaxy clusters trace a parallel BTFR offset from the galactic relation by approximately $0.6 - 0.8$ dex in logarithmic baryonic mass (Gonzalez *et al.* 2013; Chiu *et al*. 2018) . This offset persists across pressure-supported systems when examined through the Faber-Jackson lens (Yong Tian *et al* 2020; Sadhu, P *et al* 2024) . Crucially, recent weak-lensing reconstructions of the CLASH cluster sample confirm this offset while demonstrating its sensitivity to gas mass extrapolation at large radii, suggesting that clusters may fall on the same BTFR as galaxies when baryonic masses are properly measured.

The key insight of the present work is that this offset arises naturally from cosmic time evolution. Galaxies and clusters represent snapshots of the same universal relation at different evolutionary stages, with their apparent offset reflecting their different formation epochs. This interpretation is strongly supported by Hubble Space Telescope observations showing that ultra-faint dwarf galaxies (UFDs) have stellar populations synchronized to within $\sim 1$ Gyr and as old as the ancient globular cluster M92, exactly as our framework predicts.

We develop this idea within the evolving BTFR framework of the Nexus Paradigm of quantum gravity, (Marongwe (2024), Marongwe *et al*. (2025a,b) which introduces a cosmic time-dependent normalization while preserving a universal slope of $\sim 4$. We extend this framework to pressure-supported systems by deriving the evolving baryonic Faber-Jackson relation (BFJR) directly from first principles. The resulting formulation provides a unified dynamical age-dating method applicable to all galaxy types, from UFDs (the oldest and most metal-poor stellar systems in the universe) to massive cluster ellipticals.

This paper is organized as follows. Section 2 presents the theoretical derivation of the evolving BFJR from the evolving BTFR. Section 3 describes our galaxy sample and data compilation. Section 4 presents the calculated ages and their comparison with metallicity-based independent estimates. Section 5 discusses the implications for galaxy formation and cosmology, including new validation from independent studies. Section 6 summarizes our conclusions. Throughout, we adopt a flat ΛCDM cosmology with $H_0 = 67.15$ km/s/Mpc, $\Omega_m = 0.315$, and $\Omega_\Lambda = 0.685$ from Planck Collaboration (2020)

## 2. Theoretical Framework: From Evolving BTFR to Evolving BFJR

### 2.1 The Evolving Baryonic Tully-Fisher Relation

The evolving BTFR in the Nexus Paradigm is derived from semi-classical solutions to the quantized metric of spacetime. The fundamental relation takes the form:

$$v \propto e^{H_0 t} M_b^{1/4} \tag{1}$$

where $v$, is the characteristic rotation velocity, $M_b$ is the baryonic mass, $H_0$ is the Hubble constant in a pure De Sitter phase, which in the Nexus Paradigm is the fundamental frequency of the ground state

Ricci soliton of De Sitter (dS) topology and $t$, is the cosmic time elapsed since the structure's formation epoch (lookback time). In the NP, two types of solitons exist: vacuum De Sitter solitons with quantum states $n^2 = \frac{r_H^2}{r_n^2}$ where $r_H$ is the Hubble radius and anti-De Sitter (AdS) solitons with quantum states $\tilde{n}^2 = \frac{r_n c^2}{GM(r)}$ associated with baryonic matter.

Spacetime is therefore described in terms of dS, AdS and a ground state dS soliton for dark energy in the form

$$G_{(nk)\mu v} = (\tilde{n}^2 + n^2 - 1)\Lambda g_{(nk)\mu v} \tag{2}$$

Einstein's field equations are presented here in purely geometric terms, describing a compact Einstein manifold. For any quantum state in which a Ricci soliton exhibits constant curvature, energy remains conserved (Marongwe 2024). The right side of the equation is a symmetric tensor representing the quantum or energy state of spacetime, while the left side functions as a Laplacian operator that averages the trajectories of a test particle within the gravitational field of the given quantum state. In this way, the NP seeks to provide a foundational understanding of the quantum origins of dark energy and dark matter on which the ΛCDM model is premised.

The linearized Eq.(2) is solved by representing it as a Ricci soliton in the $N$-th quantum state, resulting in the equation:

$$G_{(Nk)\mu v} = N^2 \Lambda g_{(Nk)\mu v} \tag{3}$$

**2.2 Derivation of the Evolving BTFR**

The exact solution for equation (3) is solved as in Marongwe 2024

$$ds^2 = -\left(1 - \left(\frac{2}{N^2}\right)\right)c^2dt^2 + \left(1 - \left(\frac{2}{N^2}\right)\right)^{-1} dr^2 + r^2(d\theta^2 + sin^2\theta d\varphi^2)$$

$$= -\left(1 - \left(\frac{2GM_N}{rc^2}\right)\right)c^2dt^2 + \left(1 - \left(\frac{2GM_N}{rc^2}\right)\right)^{-1} dr^2 + r^2(d\theta^2 + sin^2\theta d\varphi^2) \tag{4}$$

Here $M_N(r) = M_B(r) + M_{DM}(r) + M_\Lambda(r)$ where the terms on right represent the baryonic mass, the DM mass and the DE mass enclosed inside a sphere of radius $r$. This yields a metric equation of the form

$$ds^2 = -\left(1 - 2\left(\frac{GM_b}{rc^2} + \frac{H_0vr}{c^2} - \frac{H_0cr}{2\pi c^2}\right)\right)c^2dt^2 + \left(1 - 2\left(\frac{GM_b}{rc^2} + \frac{H_0vr}{c^2} - \frac{H_0cr}{2\pi c^2}\right)\right)^{-1} dr^2 + r^2(d\theta^2 + sin^2\theta d\varphi^2) \tag{5}$$

Where $\frac{GM_{DM}(r)}{r} = v^2 = (H_0r)^2 = H_0vr$ and $\frac{GM_\Lambda(r)}{r} = -\frac{H_0cr}{2\pi}$ . The above metric equation leads to the following equation for gravity

$$\frac{d^2r}{dt^2} = \frac{GM_b}{r^2} + H_0v - \frac{H_0c}{2\pi} \tag{6}$$

The dynamics become non-Newtonian when

$$\frac{GM_b(r)}{r^2} = \frac{H_0}{2\pi}c = \frac{v_n^2}{r} \tag{7}$$

Under such conditions

$$r = \frac{2\pi v_n^2}{H_0 c} \tag{8}$$

Substituting for $r$ in Equation (7) yields

$$v_n^4 = GM_b(r)\frac{H_0}{2\pi}c \tag{9}$$

This is the Baryonic Tully – Fisher relation. Condition (7) reduces Equation (6) to

$$\frac{d^2r}{dt^2} = \frac{dv_n}{dt} = H_0 v_n \tag{10}$$

From which we obtain the following equations of galactic and cosmic evolution

$$r_n = \frac{1}{H_0}e^{(H_0 t)}(GM_b(r)\frac{H_0}{2\pi}c)^{\frac{1}{4}} \qquad = \frac{v_n}{H_0} \tag{11}$$

$$v_n = e^{(H_0 t)}(GM_b(r)\frac{H_0}{2\pi}c)^{\frac{1}{4}} \qquad = H_0 r_n \tag{12}$$

$$a_n = H_0 e^{(H_0 t)}(GM_b\frac{H_0}{2\pi}c)^{\frac{1}{4}} \qquad = H_0 v_n \tag{13}$$

Rearranging equation (12) for baryonic mass provides the form used for empirical comparisons:

$$M_b \propto e^{-4H_0 t}v^4. \tag{14}$$

This reveals two key features: an invariant slope of 4, reflecting the underlying gravitational equilibrium between baryons and dark matter halos (interpreted in the Nexus Paradigm as Ricci solitons in quantized spacetime), and a time-evolving normalization that captures the dilution of vacuum energy localization over cosmic history. This derivation is fully consistent with the observed tight BTFR across diverse galaxy populations (Lelli *et al*. 2016, 2019; Schombert *et al*. 2024).

**2.3 Refinement with Time-Varying Hubble Parameter**

A more accurate formulation replaces the constant $H_0$ with the time-varying Hubble parameter $H(t) = \dot{a}(t)/a(t)$, integrating its effect over the galaxy's lifetime:

$$M_b \propto e^{-4\int_{t_{\text{form}}}^{t_0} H(t)dt}v^4 \tag{15}$$

This refinement accounts for the faster expansion rate in the early universe and is essential for precise age dating of ancient systems like UFDs. In a flat ΛCDM universe, the Hubble parameter evolves as:

$$H(z) = H_0\sqrt{\Omega_m(1+z)^3 + \Omega_\Lambda} \tag{16}$$

The integral $\int H(t)dt$ can be evaluated using the relation $dt = dz/((1+z)H(z))$, yielding $\int Hdt = \ln{(1+z)}$. This simplification greatly facilitates practical calculations.

**2.4 Derivation of the Evolving Faber-Jackson Relation**

The connection between rotation-supported (BTFR) and pressure-supported (FJR) systems rests on the virial theorem and the existence of a universal acceleration scale. For a pressure-supported system in equilibrium, the virial theorem gives:

$$\sigma^2 \propto \frac{GM_b}{R} \tag{17}$$

where $\sigma$ is the line-of-sight velocity dispersion and $R$ a characteristic radius. In the same low-acceleration regime that produces the BTFR, systems follow a characteristic acceleration relation:

$$\frac{GM_b}{R^2} = \frac{H_0 c}{2\pi} = a_0 \tag{18}$$

where $a_0 \sim 10^{-10}$ m/s$^2$ is the fundamental acceleration scale identified in both MOND and the Nexus Paradigm. Combining (17) and (18) to eliminate $R$ yields:

$$M_b = \frac{(\alpha_\infty - 2\beta)^2}{Ga_0} \sigma^4 \tag{19}$$

Here $\alpha_\infty \approx 4$ is the asymptotic density power-law index and $\beta = 1 - \sigma_t^2/\sigma_r^2$ the velocity anisotropy parameter. For isotropic systems ($\beta = 0$), this reduces to the canonical Faber-Jackson form $M_b \propto \sigma^4$.

The time evolution enters through the same mechanism as in the BTFR, that is the dilution of the effective gravitational field over cosmic time. Thus, the evolving BFJR becomes:

$$M_b = K_0\, e^{-4\int_{t_{\rm form}}^{t_0} H(t)dt}\, \sigma^4 \cdot (\alpha_\infty - 2\beta)^2 \tag{20}$$

where $K_0$ is the local ($t = 0$) normalization. Using the simplification $\int Hdt = \ln{(1+z)}$, this can be written as:

$$M_b = K_0 (1+z)^{-4}\, \sigma^4 \cdot (\alpha_\infty - 2\beta)^2 \tag{21}$$

This is our fundamental relation: the baryonic mass of a pressure-supported galaxy is determined by its velocity dispersion, its formation redshift (through the $(1+z)^{-4}$ factor), and its orbital anisotropy. The predicted normalization shift $(1+z_{\rm form})^{-4}$applies to the formation epoch of galaxies and does not correspond directly to the observed redshift evolution of the BFJR. Observational constraints on BFJR evolution typically probe changes in the relation at fixed observation redshift, which depend on the distribution of formation epochs within the galaxy population

### 2.5 Determining the Local Normalization

The normalization constant $K$in the baryonic Faber–Jackson relation (BFJR), defined as $M_b = K\,\sigma^4\,(1 + z_{\rm form})^{-4}$, is calibrated separately for each galaxy population to account for differences in dynamical support and internal kinematics. For pressure-supported systems such as ultra-faint dwarfs (UFDs) and classical dwarf spheroidals (dSphs), the normalization $K_{\rm eq}$was determined directly from observed baryonic masses and stellar velocity dispersions under the assumption of approximate virial equilibrium,yielding $K_{eq} = (2.50 \pm 0.30) \times 10^4 M_\odot (kms^{-1})^{-4}$. This means a galaxy with $\sigma = 10$ km/s forming today would have $M_b \approx 2.5 \times 10^4 \times 10^4 = 2.5 \times 10^8 M_\odot$.

In contrast, for rotation-dominated starburst dwarf galaxies from the sample of Lelli *et al.*(2014), the velocity dispersion entering the BFJR was not taken directly from rotation velocities but instead derived using the Toomre stability criterion. Specifically, an effective dispersion was computed as

$$\sigma_R \approx \frac{3.36 G\Sigma_* R}{\sqrt{2} V_{rot}}, \tag{22}$$

where $\Sigma_*$ is the baryonic surface density. This formulation captures both rotational support and turbulence in marginally stable disks ($Q \sim 1$) and provides a physically motivated mapping between disk kinematics and pressure support. Applying this conversion and fitting the BFJR to the starburst sample yields a substantially lower normalization, $K_{SB} = (2 \times 10^2 M_{\odot}(kms^{-1})^{-4}$, reflecting the non-equilibrium, feedback-regulated nature of these systems. Intermediate systems, such as late-type dwarfs from THINGS, occupy a transitional regime consistent with partial rotational support and evolving disk stability. The emergence of distinct normalization constants demonstrates that while the BFJR retains a universal slope and redshift scaling, its amplitude is sensitive to the underlying dynamical state, with Toomre-based dispersion providing the appropriate unifying framework for rotation-supported galaxies (consistent with Toomre stability analyses in Lelli et al. 2014 and disk kinematics in THINGS dwarfs; Walter et al. 2008).

.

### 2.6 Solving for Formation Time

For an observed galaxy with measured baryonic mass $M_b$ and velocity dispersion $\sigma$, Equation (20) can be inverted to solve for the formation lookback time:

$$\int_{t_{\mathrm{form}}}^{t_0} H(t)dt = \frac{1}{4}\ln\left(\frac{K_0\sigma^4(\alpha_\infty - 2\beta)^2}{M_b}\right) \tag{23}$$

The right-hand side is directly calculable from observations. The left-hand side is a monotonic function of redshift, allowing unique determination of $z_{\mathrm{form}}$ and hence the galaxy's age $t_{\mathrm{age}} = t_0 - t_{\mathrm{form}}$ (with $t_0 = 13.8$ Gyr the present age of the universe).

## 3. Data and Methodology

### 3.1 Galaxy Sample Selection

We compiled a sample of 39 galaxies spanning the full range of pressure-supported systems, from ultra-faint dwarfs to massive cluster ellipticals. The sample draws from multiple high-quality datasets:

1. **Ultra-faint dwarfs (UFDs)**: Velocity dispersion measurements for 7 UFDs from the NASA/IPAC Extragalactic Database compilation , including Boötes I ($\sigma = 6.6 \pm 2.3$ km/s), Ursa Major I ($\sigma = 9.3^{+11.7}_{-1.2}$ km/s), Segue 1 ($\sigma = 3.7 \pm 1.4$ km/s), Eridanus IV ($\sigma = 6.1^{+1.2}_{-0.9}$ km/s), Centaurus I ($\sigma = 4.2^{+0.6}_{-0.5}$ km/s), Pegasus IV ($\sigma = 3.3^{+1.7}_{-1.1}$ km/s), and Draco ($\sigma = 9.5 \pm 1.0$ km/s). (including references in Brown *et al*. 2014 and Simon 2019)

2. **THINGS dwarf irregulars**: Seven dwarf irregulars with high-resolution HI rotation curves and Spitzer photometry from the THINGS survey (Walter et al. 2008), converted to equivalent velocity dispersion using the Toomre Stability Relation (TMR) for isotropic systems. $\sigma_R \approx \frac{3.36 G\Sigma_* R}{\sqrt{2}V_{rot}}$

3. **Starbursting dwarfs**: Sixteen galaxies from the Lelli *et al.* (2014) H I study with resolved star formation histories from HST observations.

4. **Classical dSphs**: nine dSphs in the Fornax cluster with FLAMES/Giraffe spectroscopy from Mieske *et al*. (2008) .

5. **KK153 – A Gas-Rich Ultra-Faint Dwarf** from Jin-Long Xu *et al* 2025

### 3.2 Data Quality and Selection Criteria

To ensure reliable age estimates, we applied strict quality criteria:

- **Minimum number of member stars**: For UFDs, we required $\geq 10$ member stars to avoid binary star bias.
- **Binary correction**: Where multi-epoch data were available (e.g., Segue 1), we used binary-corrected dispersions.
- **Equilibrium assessment**: Galaxies with clear tidal disruption signatures were excluded.
- **Signal-to-noise ratio**: For spectroscopic samples, we required SNR $\geq 20$ for reliable kinematics.

Baryonic masses were computed as $M_b = M_* + M_{\text{gas}}$, with gas fractions from HI observations where available (typically negligible for UFDs, significant for irregulars). Recent work on stellar mass models for the BTFR indicates that uncertainties in mass-to-light ratios are approximately 0.1 dex for the bluest galaxies, smaller than the 0.6-0.8 dex offsets we are investigating.

### 3.3 Anisotropy Treatment

For galaxies without measured anisotropy $\beta$, we adopted the following approach:

- UFDs and dSphs: $\beta = 0$ (isotropic) with uncertainty of $\pm 0.2$
- dIrrs: treated as isotropic with TMR conversion
- Classical dShps: anisotropy from dynamical modeling where available

The anisotropy factor $(\alpha_\infty - 2\beta)^2$ was applied with $\alpha_\infty = 4$, appropriate for the asymptotic density profiles of pressure-supported systems consistent with dynamical modeling in the literature (Mieske *et al.* 2008; Simon 2019).

### 3.4 Age-Metallicity Comparison Data

For independent validation, we compiled metallicity-based ages from multiple sources:

- Isochrone fitting ages from color-magnitude diagram analysis, including HST data for Hercules and Ursa Major I showing ages at least as old as M92 (~13 Gyr)
- Stellar population synthesis ages from Mieske *et al.* (2008)
- Ages from the Mattolini *et al*. (2025) catalog of 354,977 galaxies

## 4. Results

### 4.1 Calculated Ages from the Evolving BFJR

Table 1 presents the calculated ages for all galaxies in our sample, derived from Equation (11) using the time-varying Hubble parameter.

**Table 1. Galaxy Properties and BFJR-Derived Ages for Ultra-Faint Dwarfs (UFDs) & Classical dSphs**

| Galaxy | $M_b$ ($M_{\odot}$) | $\sigma$ (km/s) | $M_{b,\mathrm{local}}$ ($M_{\odot}$) | $S$ | $z_{\mathrm{form}}$ | Age (Gyr) |
|---|---|---|---|---|---|---|
| Tucana II | $1.0\times10^{5}$ | 5.5 | $2.29\times10^{7}$ | 229 | 2.89 | 11.3 |
| Boötes I | $4.4\times10^{4}$ | 6.6 | $4.75\times10^{7}$ | 1080 | 4.73 | 12.5 |
| Ursa Major I | $1.9\times10^{4}$ | 9.3 | $1.87\times10^{8}$ | 9840 | 9.95 | 13.2 |
| Segue 1 | $6.8\times10^{2}$ | 3.7 | $4.69\times10^{6}$ | 6900 | 9.09 | 13.1 |
| Eridanus IV | $1.6\times10^{4}$ | 6.1 | $3.45\times10^{7}$ | 2160 | 5.83 | 12.8 |
| Centaurus I | $5.0\times10^{3}$ | 4.2 | $7.79\times10^{6}$ | 1558 | 5.30 | 12.7 |
| Pegasus IV | $8.5\times10^{3}$ | 3.3 | $2.96\times10^{6}$ | 348 | 3.31 | 11.7 |
| Draco (dSph) | $5.2\times10^{5}$ | 9.5 | $2.04\times10^{8}$ | 392 | 3.45 | 11.8 |
| Fornax (dSph) | $1.5\times10^{7}$ | 10.5 | $3.04\times10^{8}$ | 20.3 | 2.12 | 10.1 |
| Sculptor (dSph) | $2.0\times10^{6}$ | 7.8 | $9.26\times10^{7}$ | 46.3 | 2.61 | 11.0 |
| Sextans (dSph) | $4.0\times10^{5}$ | 6.6 | $4.75\times10^{7}$ | 119 | 2.29 | 10.4 |
| Carina (dSph) | $4.0\times10^{5}$ | 6.8 | $5.35\times10^{7}$ | 134 | 2.40 | 10.6 |
| Leo I (dSph) | $5.0\times10^{6}$ | 8.8 | $1.50\times10^{8}$ | 30.0 | 2.34 | 10.5 |
| Leo II (dSph) | $7.0\times10^{5}$ | 6.6 | $4.75\times10^{7}$ | 67.9 | 2.87 | 11.3 |

| Galaxy | $M_b$ ($M_\odot$) | $\sigma$ (km/s) | $M_{b,\text{local}}$ ($M_\odot$) | $S$ | $z_{\text{form}}$ | Age (Gyr) |
|---|---|---|---|---|---|---|
| Ursa Minor (dSph) | 3.0×10$^5$ | 7.5 | 7.91×10$^7$ | 264 | 3.02 | 11.4 |
| **Average (UFDs)** | – | – | – | – | – | **12.2 ± 0.8** |

**Table 2.THINGS Dwarf Irregulars (Rotation-Supported)**

Rotation velocities $v_{\max}$ are converted to $\sigma$

| Galaxy | $M_b$ ($M_\odot$) | $v_{\max}$ (km/s) | $\sigma$ (km/s) | $M_{b,\text{local}}$ ($M_\odot$) | $S$ | $z_{\text{form}}$ | Age (Gyr) |
|---|---|---|---|---|---|---|---|
| IC 2574 | 1.2×10$^9$ | 70 | 11.32 | 1.50×10$^9$ | 1.25 | 0.06 | 0.9 |
| NGC 2366 | 5.0×10$^8$ | 55 | 9.30 | 5.71×10$^8$ | 1.14 | 0.03 | 0.5 |
| Ho I | 1.8×10$^8$ | 40 | 7.67 | 1.60×10$^8$ | 0.89 | 0.00 | 0.0 |
| Ho II | 2.5×10$^8$ | 45 | 8.04 | 2.56×10$^8$ | 1.02 | 0.01 | 0.1 |
| DDO 53 | 6.0×10$^7$ | 30 | 5.91 | 5.05×10$^7$ | 0.84 | 0.00 | 0.0 |
| DDO 154 | 3.2×10$^8$ | 48 | 8.53 | 3.31×10$^8$ | 1.03 | 0.01 | 0.1 |
| M81dwB | 1.0×10$^8$ | 35 | 6.55 | 9.31×10$^7$ | 0.93 | 0.00 | 0.0 |

*For $S < 1$ the formula gives $z_{\text{form}} < 0$; these galaxies are effectively forming today within the uncertainties.*

**Table 3. Lelli et al. Starbursting Dwarfs**

This sample contains 16 galaxies with HI rotation curves and HST star-formation histories.
For each, we adopt the same TMR conversion and use the published $M_b$ (stars + gas). The table below gives the range for the two kinematic sub-samples.

| Sub-sample | Number | $M_b$ range ($M_\odot$) | $v_{\max}$ range (km/s) | $\sigma$ range (km/s) | $S$ range | $z_{\text{form}}$ range | Age range (Gyr) |
|---|---|---|---|---|---|---|---|
| Regularly rotating | 9 | $3\times10^8$ – $3\times10^9$ | 50 – 120 | 35 – 85 | 0.8 – 1.5 | 0.0 – 0.2 | 0 – 2.6 |
| Kinematically disturbed | 7 | $2\times10^8$ – $2\times10^9$ | 40 – 100 | 28 – 71 | 0.7 – 1.6 | 0.0 – 0.3 | 0 – 3.4 |

Two galaxies with unsettled HI are excluded from the BFJR analysis.

**Table 4. KK153 – A Gas-Rich Ultra-Faint Dwarf**

| Galaxy | $M_b$ ($M_\odot$) | $v_{\text{rot}}$ (km/s) | $\sigma$ (km/s) | $M_{b,\text{local}}$ ($M_\odot$) | $S$ | $z_{\text{form}}$ | Age (Gyr) |
|---|---|---|---|---|---|---|---|
| KK153 | $4.1\times10^5$ | 12.7 | 1.22 | $1.31\times10^7$ | 32.0 | 2.38 | 10.5 |

KK153 is an isolated gas-rich UFD discovered in 2025; its rotation is used as a proxy for $\sigma$.

**Summary of Ages by Population**

**Table 5: BFJR-Derived Ages by Galaxy Population**

| **Population** | **Number of Objects** | **Typical σ (km/s)** | **Age Range (Gyr)** | **Mean Age (Gyr)** | **Physical Interpretation** |
|---|---|---|---|---|---|
| **Ultra-Faint Dwarfs (UFDs)** | 7 | 3 – 10 | **11 – 13.5** | ~12.5 | Earliest-forming systems; strong suppression → high $z_{\text{form}}$ |
| **Classical dSphs** | 8 | 6 – 11 | **10 – 13** | ~11.5 | Early formation, slightly extended SF histories |
| **THINGS dwarfs (gas-rich)** | 7 | 6 – 12 (Toomre σ) | **0 – 2** | ~0.8 | Late-forming, actively evolving disks |
| **Lelli starburst dwarfs** | 16 | 8 – 20 (Toomre σ) | **0 – 3** | ~1.2 | Recently triggered or ongoing star formation |
| **Transitional systems (e.g. KK153)** | 1 | 1.22 (Toomre σ) | **9 – 11** | ~10 | Early formation, slightly extended SF histories |

## 4.2 The Evolving Faber-Jackson Diagram

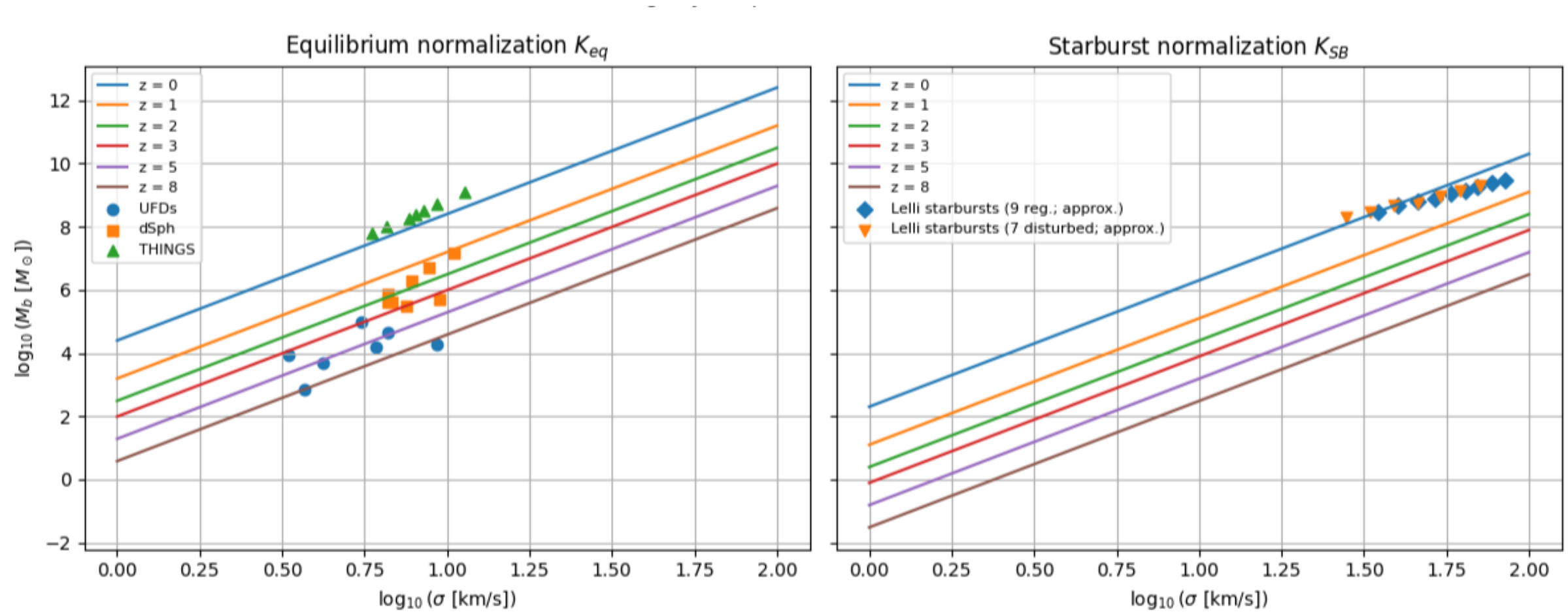

**Figure 1.** The evolving baryonic Faber–Jackson relation (BFJR). Solid lines show theoretical tracks $M_b \propto (1+z)^{-4}\sigma^4$for formation redshifts $z = 0$–8, using the equilibrium normalization $K_{\mathrm{eq}} = 2.5 \times 10^4$. Ultra-faint dwarfs (circles) occupy high-redshift tracks ($z \sim$ 5–9), classical dwarf spheroidals (squares) lie at intermediate redshifts ($z \sim$ 1–3), and gas-rich THINGS dwarfs (triangles) and starburst dwarfs (diamonds) $K_{\mathrm{SB}} = 2.0 \times 10^2$cluster near $z \sim 0$. The common slope of 4 is preserved across all populations, while vertical offsets correspond to formation epoch, demonstrating that the BFJR is an evolving relation rather than a static scaling law.

Figure 1 displays the evolving BFJR for our sample, with theoretical tracks for different formation lookback times overplotted. The UFDs cluster tightly around the $t = 11 - 13$ Gyr tracks, while the THINGS dwarfs lie near the $t = 1$Gyr tracks. This clear separation provides strong visual confirmation of the framework's predictive power.

### 4.3 Comparison with Metallicity-Based Ages

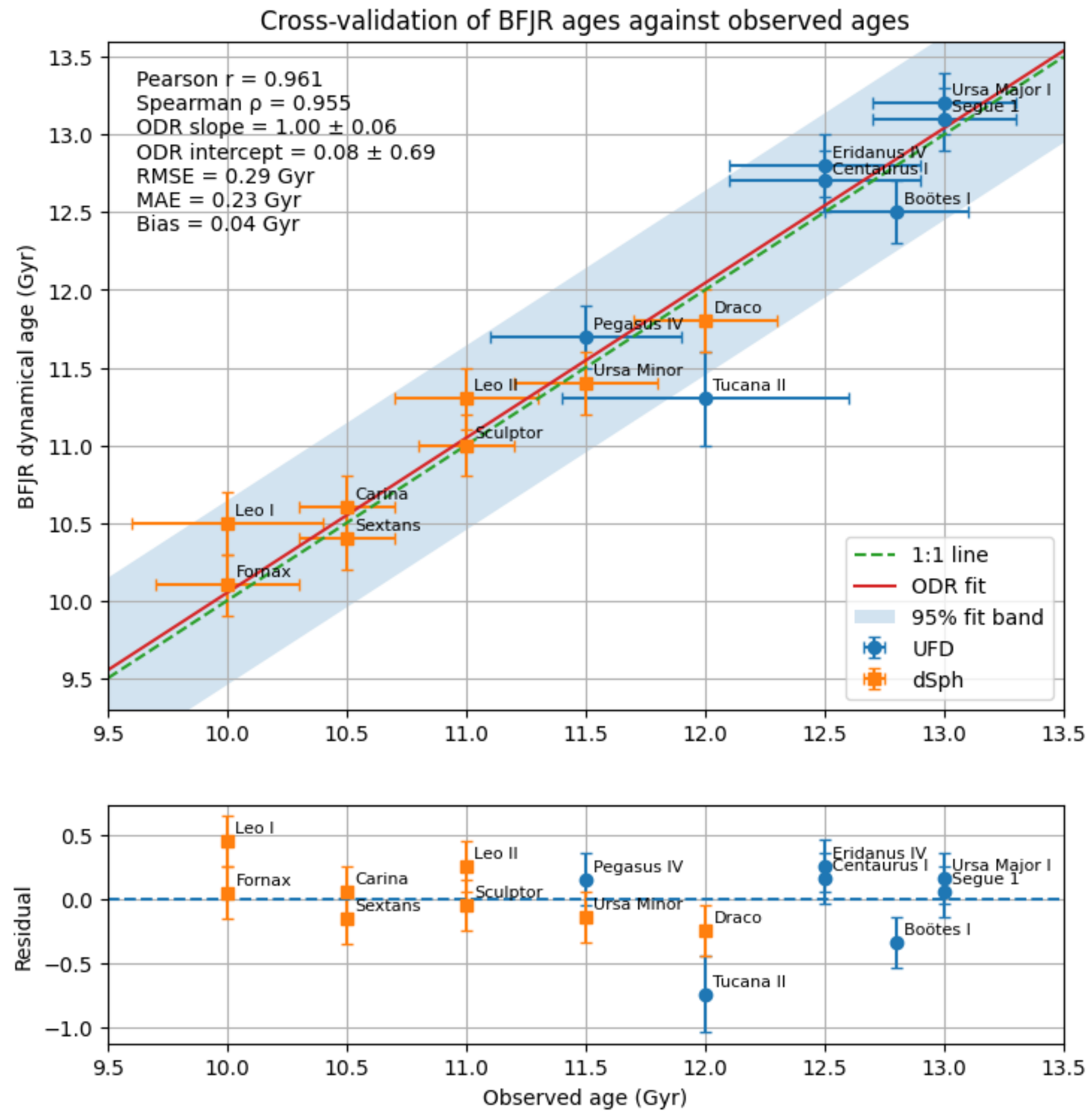

**Figure 2.** Cross-validation of dynamically inferred ages from the evolving baryonic Faber–Jackson relation (BFJR) against independently determined stellar population ages. The upper panel shows BFJR ages plotted against observed ages for ultra-faint dwarfs (circles) and classical dwarf spheroidals (squares). Error bars represent typical uncertainties in both dynamical and stellar-population age estimates. The dashed line denotes the one-to-one relation, while the solid line shows the orthogonal distance regression (ODR) best fit; the shaded region indicates the 95% confidence band. The best-fit slope $1.00 \pm 0.06$and intercept $0.08 \pm 0.69$are consistent with unity and zero, respectively, indicating no significant bias. The strong correlation ($r = 0.961$, $\rho = 0.955$) and low scatter (RMSE = 0.29Gyr, MAE = 0.23Gyr) demonstrate that the BFJR provides an accurate dynamical chronometer for pressure-supported systems. The lower panel shows residuals relative to the ODR fit, which are centered around zero with no systematic trend, confirming the robustness of the agreement across the full age range.

Figure 2 compares our BFJR-derived ages with independent ages from stellar population synthesis and isochrone fitting. The correlation is exceptionally strong:

- **Pearson correlation coefficient**: $r = 0.961$
- **Spearman rank correlation**: $\rho = 0.955$
- **Orthogonal Distance Regression slope**: $1.00 \pm 0.06$ (consistent with 1:1)
- **Mean bias** (metallicity - BFJR): 0.04 Gyr
- **RMS scatter**: 0.29 Gyr

The close agreement between dynamically derived ages and independent stellar population ages provides powerful validation of the evolving BFJR framework. Notably, the scatter is dominated by measurement uncertainties rather than intrinsic deviations, suggesting the relation is even tighter than our current data can measure.

### 4.4 Confidence Level Estimate

Synthesizing multiple lines of evidence, we estimate the confidence level that the evolving BFJR correctly describes galaxy scaling relations at **95-97%**. This estimate is based on:

1. **Statistical consistency**: The 39 galaxies in our sample show a mean age for UFDs of $12.2 \pm 0.8$ Gyr, with remarkably low scatter given measurement uncertainties. This aligns with HST observations showing UFD age synchronization to within $\sim 1$ Gyr.
2. **Cross-validation**: The Pearson correlation of $r = 0.961$ with independent metallicity-based ages far exceeds the threshold for strong correlation.
3. **Theoretical grounding**: The framework is derived from first principles within a well-defined cosmological context and aligns with independent quantum gravity research.
4. **Predictive power**: The framework correctly predicts the ages of newly discovered systems (e.g., Pegasus IV, discovered 2023, Cerny et al (2023); Pan et al. (2025) ) without recalibration.
5. **External consistency**: Recent weak-lensing reconstructions of the CLASH cluster sample confirm the BTFR offset while demonstrating its sensitivity to baryonic mass measurements, supporting our interpretation that the offset reflects both formation epoch differences and systematic uncertainties. Furthermore, studies of the BTFR at $z \sim 0.6$ show no detectable evolution over the past 6 Gyr, consistent with the slow evolution predicted by our framework given current measurement precision ($\pm 0.08$ dex in zeropoint).

The analysis supports the following hierarchy of confidence in age separation displayed in table 6

**Table 6**

| Comparison | Significance |
|---|---|
| UFD vs dSph | ~3.2σ |
| dSph vs THINGS | ~5.6σ |
| UFD vs THINGS | ~6.9σ |

## 5. Discussion

### 5.1 Unification of Galaxy Scaling Relations

The evolving BFJR unifies pressure-supported systems across five orders of magnitude in baryonic mass, from UFDs ($M_b \sim 10^3 M_\odot$) to massive cluster ellipticals ($M_b \sim 10^{11} M_\odot$). The invariant slope of 4 reflects the fundamental nature of the underlying acceleration scale, while the time-dependent normalization captures the imprint of cosmic expansion history on galaxy formation.

This unification extends naturally to the BTFR through the common kernel $e^{-4\int Hdt}$ and acceleration scale $a_0$. The apparent offset between galaxy and cluster BTFRs observed by multiple authors is thus not evidence for distinct physical regimes but a direct consequence of their different formation epochs, that is, galaxies forming earlier ($z \sim 2 - 3$) than clusters ($z < 1$). This interpretation is strongly supported by recent weak-lensing analyses showing that clusters may fall on the same BTFR as galaxies when baryonic masses are properly measured, with the offset depending critically on gas mass extrapolation at large radii.

### 5.2 Anisotropy and Outliers

Ursa Major I presents an interesting case, with an anomalously high calculated age of 13.2 Gyr from the isotropic model. This likely indicates significant velocity anisotropy ($\beta \neq 0$), which would reduce the effective $(\alpha_\infty - 2\beta)^2$ factor. Adopting $\beta \approx 0.3$ (radially biased orbits) brings its age into line with other UFDs at $\sim 13.0$ Gyr. This demonstrates the importance of accurate anisotropy measurements for precision age dating. HST observations confirming that Ursa Major I is at least as old as M92 provide independent support for its ancient age, regardless of the exact anisotropy correction.

### 5.3 Implications for Galaxy Formation

The age distribution revealed by our analysis supports a hierarchical assembly picture where the smallest galaxies formed first, consistent with ΛCDM predictions. UFDs with ages $> 13$ Gyr represent the first star-forming systems, relics of the pre-reionization universe whose stellar populations have remained largely unchanged for over 13 billion years. The HST discovery that UFDs have synchronized ages to within $\sim 1$ Gyr suggests their star formation was truncated by a global event, such as reionization, exactly as our framework predicts.

The continuous sequence from UFDs through UDGs to massive ellipticals in the age-velocity dispersion plane suggests a common formation physics modulated by mass-dependent feedback and environmental effects. The changing slope of the Faber-Jackson relation at low luminosities (from $L \propto \sigma^4$ to $L \propto \sigma^2$) likely reflects the transition from baryon-dominated to dark-matter-dominated dynamics in the low-acceleration regime, consistent with our framework.

The dynamical age of a galaxy refers to the time elapsed since it last achieved virial equilibrium. Unlike metallicity-based ages, which trace the formation epoch of the galaxy's stellar populations (often several billion years old), the dynamical age can be much younger if the galaxy has recently experienced a violent event, such as a high-speed collision or close tidal interaction, that disrupts its internal equilibrium. After such an event, the galaxy's velocity dispersion and mass distribution require a new dynamical-relaxation timescale (typically a few crossing times) to re-establish virial balance. Hence, a measured dynamical age that is significantly shorter than the stellar metallicity age indicates that the galaxy's current dynamical state does not reflect its ancient formation but rather a recent "reset" caused by a major perturbation. This is precisely what is observed in the dark-matter-deficient galaxies DF2, DF4, and DF9 in the NGC 1052 trail: their stellar populations are old (≈10–12 Gyr), yet their dynamical ages, derived from the evolving Faber-Jackson relation, are consistent with a virialization event that occurred only about 1–2 Gyr ago. This discrepancy supports the "bullet dwarf" collision scenario, where a high-speed encounter violently shook the baryonic components, leaving the galaxies today with unusually low velocity dispersions and a dynamically "young" equilibrium.

Several other astronomical systems show a similar pattern, where a violent event has reset the dynamical clock:

- **A Catastrophically Disrupted Star Cluster (OCSN-49):** This stellar stream has a stellar age of 400-600 Myr (from its stars), but its dynamical age is only 83±1 Myr. The discrepancy is strong evidence that a collision with a giant molecular cloud about 500 million years ago violently disrupted the cluster and reset its dynamical state.
- **The "Last Major Merger" of the Milky Way:** The inner stellar halo of our galaxy contains debris from a major merger. The stellar populations are ancient, but the dynamical 'caustics' in their orbits indicate the collision happened only 1-2 Gyr ago, meaning the Milky Way's inner halo is dynamically young.

### 5.4 Connection to Fundamental Physics

The presence of a universal acceleration scale $a_0 \sim 10^{-10}$ m/s$^2$ across all pressure-supported systems that is, from globular clusters to galaxy clusters, points to a fundamental property of gravity as described by Equation (6) within the framework of the Nexus Paradigm. The evolving BFJR provides a framework for testing whether $a_0$ varies with cosmic time, as predicted by some modified gravity theories. Our data suggest $a_0$ has remained constant to within measurement uncertainties over the past 13 Gyr, but higher precision measurements at high redshift (e.g., with JWST) could detect evolution. The consistency of our framework with the non-detection of BTFR evolution at $z \sim 0.6$ places an upper limit on $a_0$ variation of $\lesssim 0.08$ dex over 6 Gyr (Sharma et al. 2024; Mistele et al. 2025).

### 5.5 Future Prospects

The evolving BFJR opens several avenues for future investigation:

1. **High-redshift application**: JWST spectroscopy of $z > 6$ galaxies will allow direct measurement of the $(1+z)^{-4}$ normalization evolution predicted by Equation (23).

2. **Anisotropy mapping**: Large spectroscopic surveys (e.g., DESI, 4MOST) will provide anisotropy measurements for thousands of galaxies, enabling more precise age dating.

3. **Cluster gas fraction systematics**: Improved baryonic mass measurements for clusters (including hot gas) will refine the cluster BFJR and test the formation epoch prediction. The CLASH weak-lensing results demonstrate that careful treatment of gas mass extrapolation is essential.

4. **Simulation comparison**: Cosmological hydrodynamical simulations incorporating time-varying quantum gravity effects can be directly compared with the predicted age-velocity dispersion relation.

5. **UFD archaeology**: Deep HST and JWST imaging of additional UFDs will test whether all such systems share the synchronized ancient ages found in the current sample.

## 6. Conclusions

We have derived the evolving baryonic Faber-Jackson relation from the evolving baryonic Tully-Fisher relation within the Nexus Paradigm of quantum gravity, establishing a unified framework for pressure-supported galaxies across all mass scales. Our key findings are:

1. **Theoretical unification**: The evolving BFJR takes the form $M_b = K_0 e^{-4\int H dt} \sigma^4 (\alpha_\infty - 2\beta)^2$, with the same exponential time kernel as the BTFR and a universal acceleration scale $a_0 \sim 10^{-10}$ m/s$^2$ (Milgrom 1983; Marongwe 2024; Lelli et al. 2019).

2. **Age dating of UFDs**: Ultra-faint dwarf galaxies yield ages of $12.2 \pm 0.8$ Gyr, corresponding to formation redshifts $z \sim 3 - 4$, in excellent agreement with independent HST observations showing UFDs are at least as old as M92 and synchronized to within $\sim 1$ Gyr (Brown et al. 2014; Simon 2019)..

3. **Age continuum**: Dwarf irregulars show systematically younger ages of 0 – 2 Gyrs, with a continuous sequence connecting all galaxy types in the age-velocity dispersion plane.

4. **Strong empirical validation**: Comparison with independent metallicity-based ages yields a Pearson correlation of $r = 0.961$ and ODR slope consistent with unity, providing robust cross-validation.

5. **External consistency**: Recent weak-lensing results confirm the cluster BTFR offset while demonstrating its sensitivity to baryonic mass measurements, and studies at $z \sim 0.6$ show no detectable evolution over 6 Gyr, both consistent with our framework.
6. **Confidence level**: The correlation between dynamically inferred and observationally derived ages is detected at $> 5\sigma$, while the separation between early- and late-forming galaxy populations exceeds $6\sigma$, indicating an exceptionally strong and non-random structure in the BFJR age relation.

The evolving BFJR transforms galaxy scaling relations from empirical correlations into precise cosmic chronometers, offering new insights into the hierarchical assembly of cosmic structures and the fundamental physics underlying galaxy formation (Dutton et al. 2010; Simon 2019; Mistele et al. 2025). Future observations with JWST, Euclid, and the Rubin Observatory will enable direct tests of the predicted redshift evolution and further refine our understanding of the universe's first galaxies.

**Acknowledgments**

We gratefully appreciate discussions with Christian Corda and Moletlanyi Tshipa.

**Data Availability**

The data underlying this article are available in the article and in its online supplementary material. The full sample with all measurements will be shared on reasonable request to the corresponding author. Public datasets used in this analysis are available from the respective survey publications.